\documentclass[twocolumn,amsmath,amssymb]{revtex4-1}
\usepackage{graphicx}
\usepackage{dcolumn}
\usepackage{bm}
\usepackage{amsmath}
\usepackage{physics}

\let\oldAA\AA
\renewcommand{\AA}{\text{\normalfont\oldAA}}

\usepackage{amsmath}
\usepackage{subfigure}
\usepackage{hyperref}
\usepackage[normalem]{ulem}
\hypersetup{
    colorlinks,
    citecolor=blue,
    filecolor=black,
    linkcolor=blue,
    urlcolor=blue
}
\usepackage{epstopdf}
 \usepackage{relsize}

\newcommand*{\rom}[1]{\expandafter\@slowromancap\romannumeral #1@}
\title{Report}

\setcitestyle{square}

\begin{document}


\title{Realistic non-local refrigeration engine based on Coulomb coupled systems }

\author{Anamika Barman}
\affiliation{%
Department of Electronics and Electrical Communication Engineering,\\
Indian Institute of Technology Kharagpur, Kharagpur-721302, India\\
}%

\author{Surojit Halder}
\affiliation{%
	Department of Electronics and Electrical Communication Engineering,\\
	Indian Institute of Technology Kharagpur, Kharagpur-721302, India\\
}%

\author{Shailendra K. Varshney
}
\affiliation{%
	Department of Electronics and Electrical Communication Engineering,\\
	Indian Institute of Technology Kharagpur, Kharagpur-721302, India\\
}%

\author{Gourab Dutta
}
\affiliation{%
	Department of Electronics and Electrical Communication Engineering,\\
	Indian Institute of Technology Kharagpur, Kharagpur-721302, India\\
}%

\author{Aniket Singha
}
\affiliation{%
	Department of Electronics and Electrical Communication Engineering,\\
	Indian Institute of Technology Kharagpur, Kharagpur-721302, India\\
}%




\begin{abstract}
Employing Coulomb-coupled systems, we demonstrate a cryogenic non-local refrigeration engine, that circumvents the need for a change in the energy resolved system-to-reservoir coupling, demanded by the recently proposed non-local refrigerators \cite{coulomb_TE1,coulomb_TE2,coulomb_TE3,coulomb_TE4,coulomb_TE5}.   We demonstrate that  an intentionally introduced   energy difference  between the  ground states of adjacent tunnel coupled quantum dots, associated with Coulomb coupling, is sufficient  to extract heat from a remote target reservoir. 
Investigating the performance and operating regime   using quantum-master-equation (QME) approach, we point out to some crucial aspects of the proposed refrigeration engine.  In particular,  we demonstrate that the maximum cooling power  for the proposed set-up is limited to about $70\%$ of the optimal design. Proceeding further, we point out that to achieve a target reservoir temperature, lower compared to the average temperature of the current path, the applied voltage must be greater than a given threshold voltage $V_{TH}$, that increases with decrease in the target reservoir temperature. In addition, we demonstrate that the maximum cooling power, as well as the coefficient of performance deteriorates as one approaches a lower target reservoir temperature.  The novelty of the proposed refrigeration engine  is  the integration of fabrication simplicity along with descent cooling power. The idea proposed in this paper may   pave the way towards the realization of efficient non-local cryogenic refrigeration systems. 

\end{abstract}
\maketitle

\section{Introduction}
\begin{figure}
	\includegraphics[width=.5\textwidth]{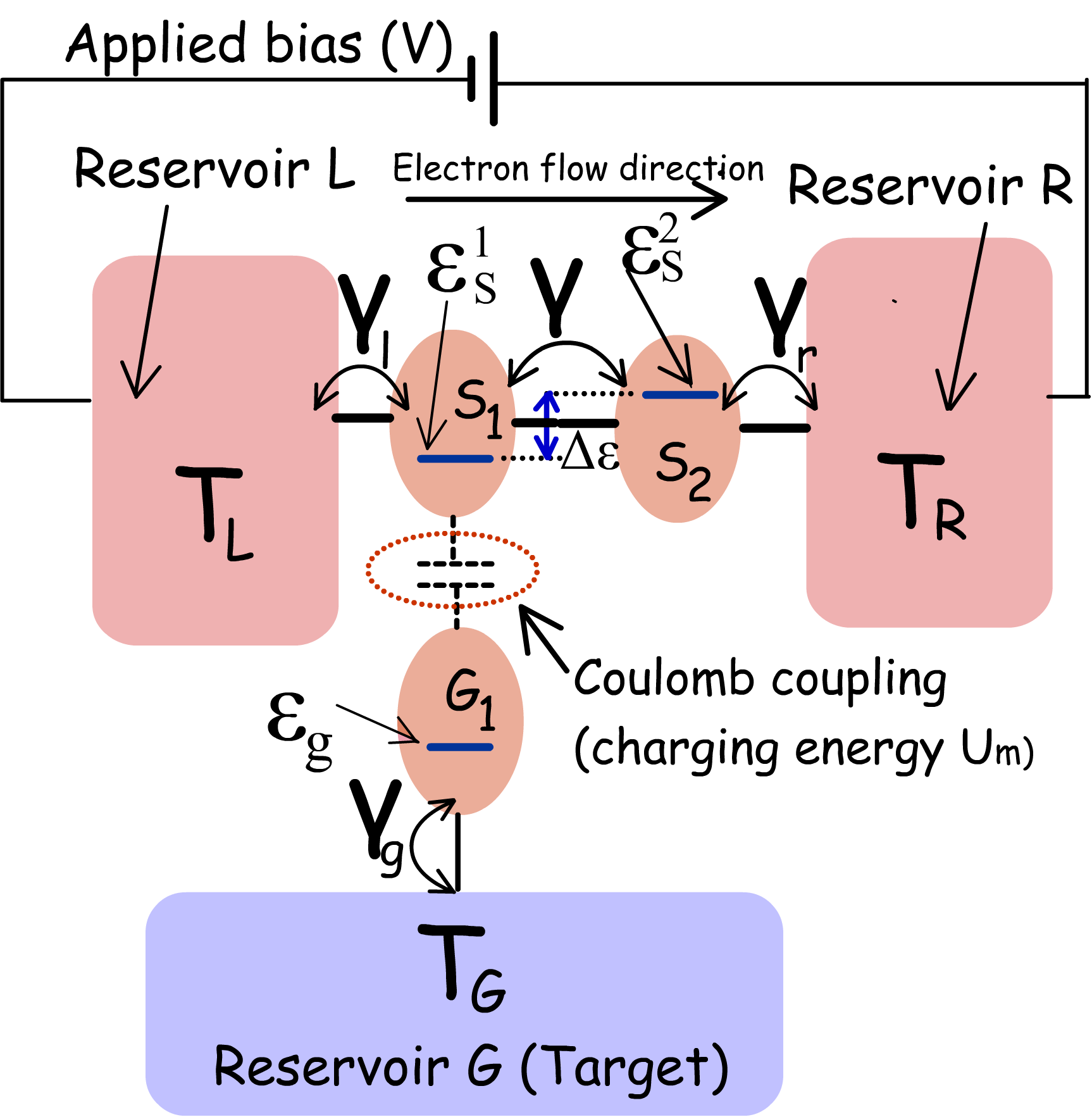}
	\caption{Schematic of the proposed non-local refrigeration engine employing Coulomb coupled systems. The system consists of three dots-$S_1,~S_2$ and $G_1$, which are electrically coupled to the macroscopic electron reservoirs $L,~R$ and $G$ respectively.  $S_1$ is tunnel coupled to $S_2$ and capacitively coupled to $G_1$.  The tunnel coupled quantum dots $S_1$ and $S_2$ share a stair-case ground state configuration with $\varepsilon_s^2=\varepsilon_s^1+\Delta\varepsilon$.}
	\label{fig:Fig_1}
\end{figure}
With  scaling technology rapidly invading the nano-domain, the tremendous rise in dissipated heat density and hence operating temperature has drawn significant attention towards electrical refrigeration in nano-scale dimensions \cite{coulomb_TE1,coulomb_TE2,coulomb_TE3,coulomb_TE4,coulomb_TE5,snyder_thompson,cooling_ref1,cooling_ref2,cooling_ref3,cooling_ref4,cooling_ref5,cooling_ref6,cooling_ref7,cooling_ref8,cooling_ref9,cooling_ref10,cooling_ref11,whitney,whitney2,aniket_cool1,kim_conserve,aniket_cool2}. In addition, sophisticated experiments on exploratory technologies, such as quantum computation, spin and optics based computation, etc. occasionally call for electrical  refrigeration at cryogenic temperatures in nano-meter range length scale.  However, the  refrigeration performance in such nano-scale systems is often affected drastically by the large lattice heat flux, particularly when both the region of refrigeration and heat dissipation lie along the path of current flow and are separated by a few nanometres in space. Despite lots of effort to reduce lattice thermal conductance \cite{phonon1,phonon2,phonon3,phonon4,phonon5,phonon6,phonon7,superlattice1,superlattice2,nanoflake_heat,nanowire_heat1,nanowire_heat2}, the performance of nano  refrigeration systems is still affected by rapid reverse heat flux. This effect poses a threat to the refrigeration performance as device channels are gradually invading the nano-domain. An attempt to  improve the  refrigeration performance by engineering lattice thermal conductance generally deteriorates the electronic conductivity  and hence the cooling power. As an alternative, one of the major research focus, concerning nano-scale refrigeration engine, is to facilitate an independent manipulation of  the  electron transport path and lattice heat  conduction path, by introducing a spatial separation between the current path and the target reservoir  \cite{sep1,sep2,sep3,coulomb_TE1,coulomb_TE2,coulomb_TE3,coulomb_TE4,coulomb_TE5}.   This phenomenon of refrigerating a remote target reservoir, which is spatially separated from the current track, is known as non-local refrigeration \cite{sep1,sep2,sep3,coulomb_TE1,coulomb_TE2,coulomb_TE3,coulomb_TE4,coulomb_TE5}. Thus, non-local refrigeration systems are three terminal systems where input power is delivered between two terminals to extract heat from a remote target reservoir through the third terminal.  In this case,  optimizing the  lattice heat transport path, in an attempt to  improve refrigeration performance, can be accomplished without modifying the current conduction path.  These kind of systems, thus enables   an independent optimization of the lattice thermal conductance  and the current conductivity \cite{sep1,sep2,sep3,coulomb_TE1,coulomb_TE2,coulomb_TE3,coulomb_TE4,coulomb_TE5}. In addition, due to the non-locality of  the electronic transport path,  the refrigerated region is significantly shielded from reverse heat flux owing to the Joule dissipation. \\   
\indent Recently, optimal non-local refrigeration systems based on Coulomb coupled quantum dots have been proposed and explored in the literature \cite{coulomb_TE1,coulomb_TE2,coulomb_TE3,coulomb_TE4,coulomb_TE5}.  However, the operation of these refrigeration engines call for a sharp step-like change in the energy-resolved system-to-reservoir coupling around the quantum-dot ground state  \cite{coulomb_TE1,coulomb_TE2,coulomb_TE3,coulomb_TE4,coulomb_TE5}, which is impossible to achieve in realistic systems. In this paper, we propose a design strategy for non-local refrigeration using Coulomb coupled systems, that can operate optimally without demanding a sharp transition in the system-to-reservoir coupling.  The refrigeration engine is then theoretically analyzed using  quantum master equation (QME) approach for such systems   in the sequential tunneling limit \cite{sispad}.  
It is demonstrated that the maximum cooling power (heat extracted per unit time) for the proposed design is limited to  about $70\%$ of that for the optimal design \cite{coulomb_TE1,coulomb_TE2,coulomb_TE3,coulomb_TE4,coulomb_TE5}. Despite of a lower performance compared to the optimal set-up, the novelty of  the proposed design lies in the integration of  fabrication simplicity  along with a decent cooling power, making such design suitable for practical applications. At the end, the sequential transport phenomena leading to a deterioration of performance of the proposed set-up is investigated and a strategy to alleviate such transport processes is presented.\\
\indent The paper is organized as follows. In Sec.~\ref{design}, we briefly describe the proposed design along with the transport formulation employed to analyze the performance of the refrigeration engine. In the following Sec.~\ref{results}, we investigate the performance and  region of operation of the proposed refrigeration engine for two different cases (i)  $T_G<T_{L(R)}$ and  $T_G=T_{L(R)}$. This section also presents a performance comparison of the proposed refrigeration engine with the optimal set-up proposed in literature \cite{coulomb_TE1,coulomb_TE2,coulomb_TE3,coulomb_TE4,coulomb_TE5}, in addition to an investigation of the sequential processes leading to a performance deterioration of the proposed set-up. Finally, we conclude this paper briefly in Sec.~\ref{conclusion}.


\section{Proposed design and transport formulation}\label{design}

The proposed non-local refrigeration engine is schematically illustrated in Fig.~\ref{fig:Fig_1}, where  three quantum dots $S_1,~ S_2$ and  $G_1$  are electrically coupled with electronic reservoirs $L, ~R$ and $ G$ respectively, $G$ being the target reservoir to be refrigerated. $S_1$ and  $S_2$ are tunnel coupled to each other, while  $G_1$ is capacitively coupled to $S_1$ by suitable fabrication techniques.  The capacitive coupling between $S_1$ and $G_1$  permits energy exchange  while obstructing any particle swap between the dots.  In the optimal Coulomb-coupled system based refrigeration-engine, an asymmetric system-to-reservoir coupling is required for refrigeration \cite{coulomb_TE1,coulomb_TE5}.  In the proposed set-up, on the other hand,  the asymmetricity, with respect to the reservoir $L$ and $R$, is embedded within the system itself by choosing a difference between the ground states of  the tunnel-coupled quantum dots $S_1$ and $S_2$, with $\varepsilon_s^2=\varepsilon_{s}^1+\Delta \varepsilon$. The electrically coupled dots $S_1$ and $S_2$  may be suitably fabricated or gated to retain a stair-case ground-state configuration with  $\varepsilon_s^2=\varepsilon_s^1+\Delta\varepsilon$.    we will demonstrate via numerical calculations and theoretical arguments  that in the system detailed above,  refrigeration of the target reservoir $G$  can be achieved  by forcing a net electronic flow from $L$ to $R$, that is, refrigeration can be achieved at a terminal  non-local to the current path.  The excess energy   $\Delta \varepsilon=\varepsilon_s^2-\varepsilon_s^1$, required for the electrons to tunnel from $S_1$ to $S_2$ is extracted from the reservoir $G$ via Coulomb coupling.  Coming to the fabrication feasibility of such a system, due to the recent progress in in  nano-fabrication techniques, coupled  systems employing multiple (more than two) quantum dots, with and without Coulomb coupling, have already been  experimentally realized \cite{mqd1,mqd2,mqd3,mqd4,mqd5,mqd6}. In addition, it has been experimentally demonstrated that spatially and electrically isolated quantum dots may be bridged to obtain strong Coulomb coupling, in addition to excellent thermal insulation \cite{cap_coup_1,cap_coup_2,cap_coup3,cap_coup_4,cap_coup_5}. In addition, the bridge may be constructed between two specific  dots to radically increase their mutual Coulomb coupling, without affecting the electrostatic potential of the other quantum dots \cite{cap_coup_1,cap_coup_2,cap_coup3,cap_coup_4,cap_coup_5}.  Thus, the fluctuation in electron number $n_{S_1}~(n_{G_1})$ of the dot $S_1$ ($G_1$)  alters the electrostatic energy of the  dot $G_1$ ($S_1$).  
   The total increase in electrostatic energy $U$ of the configuration, consisting  of three dots, due to fluctuation in electron number  can be given by \cite{aniket_nonlocal,sispad}: 
	\small
	\begin{multline}
	U(n_{S_{1}},n_{G_{1}},n_{S_{2}})=\sum_{x }U^{self}_{x}\left(n_{x}^{tot}-n_x^{eq}\right)^2 \nonumber \\  +\sum_{(x_{1},x_{2})}^{x_1 \neq x_2} U^m_{x_1,x_2}\left(n_{x_1}^{tot}-n_{x_1}^{eq}\right)\left(n_{x_2}^{tot}-n_{x_2}^{eq}\right) 
	\end{multline}
\normalsize
where $n_x^{tot}$ is the total electron number at finite temperature, and $n_x^{eq}$ is the total electron number under equilibrium conditions at $0K$ in the dot $x$ (satisfying the condition that the system equilibriates at the minimum possible value of electrostatic energy at $0K$). $U_x^{self} =\frac{q^2}{C_x^{self}}$ is the electrostatic energy due to self-capacitance $C_x^{self}$ (with the adjacent terminals) of quantum dot ‘$x$’ and $U_{x_1,x_2}^m$ is the mutual electrostatic energy between two spatially separated  quantum dots $x_1$ and $x_2$. $n_x=n_x^{tot}-n_x^{eq}$ is the number of excess electrons in the ground state of dot $x$ due to thermal fluctuations (kicks) from the reservoirs at finite temperature. To investigate the performance of the refrigeration engine,  we  consider a minimal physics-based model to simplify our calculations. We assume that the electrostatic energy due to self-capacitance is much greater than the applied voltage $V$ or the average thermal voltage $kT/q$, i.e. $U_x^{self}>>(kT, qV)$, where $T=\frac{T_{L(R)}+T_G}{2}$, such that electron occupancy probability or transfer rate across the Coulomb blocked energy level due to self-capacitance is negligibly small. Hence, the ground state  of a particular dot can be occupied by atmost one electron and the behaviour of the entire system can be analyzed via $2^3=8$ different multi-electron states. These states may be  denoted as $|n_{S_1},n_{G_1},n_{S_2}\rangle=|n_{S_1}\rangle\otimes|n_{G_1}\rangle\otimes|n_{S_2}\rangle$ where $n_{S_1},~n_{G_1},~n_{S_2} \in (0,1)$ indicate the number of electrons in the ground state of  $S_1,~G_1$ and $S_2$ respectively. 
We consider that the Coulomb coupling  between $S_1-S_2$ and $S_2-G_1$ is negligible compared to the relevant energy scales of the system, that is, electrostatic coupling between $S_1$, $S_2$, and $S_2$, $G_1$ is negligible with respect  to $U^m_{S_1,G_1}, ~kT$ and $qV$.  Thus, for all practical purpose relating to electron transport $U^m_{S_1,S_2} \approx 0$ and $U^m_{G_1,S_2} \approx 0$. Due to capacitive coupling,   the electronic transport trough $S_1$ and $G_1$ are interdependent,  and hence, the  pair of dots $S_1$ and $G_1$ are treated as a sub-system ($\varsigma_1$) of the entire system, $S_2$ being the complementary sub-system labeled as $\varsigma_2$ \cite{sispad}. The  probability of the sub-system ($\varsigma_1$) to be in a particular state  is denoted by $P_{i,j}^{\varsigma_1}$,  where $i$ and $ j$ are the number of electrons in the ground state of the dots $S_1$ and $G_1$ respectively. On the other hand, $P_k^{\varsigma_2}$ denotes the steady-state   occupancy probability of dot $S_2$  (subsystem $\varsigma_2$). Under the condition that $\Delta\varepsilon$ is much higher than the ground state broadening due to reservoir coupling, the optimal  
inter-dot tunneling  between $S_1$ and $S_2$ occurs when $\Delta \varepsilon=U^m_{S_1,G_1}$, such that $\varepsilon_s^2=(\varepsilon_s^1+U^m_{S_1,G_1})$ \cite{sispad}. Hence, for the optimal  performance investigation of the proposed set-up, we assume $\Delta \varepsilon=U^m_{S_1,G_1}$. In what follows, we simply refer to $U^m_{S_1,G_1}$  as $U_m$ to make the notations compact. With all the above assumptions,  the equations dictating the steady state  sub-system  probabilities  can be obtained  as follows \cite{sispad}:
\small
\begin{widetext}
	\vspace{-.3cm}	\begin{align}
	&  -P_{0,0}^{\varsigma_1}\{f_L(\varepsilon_s^1)+f_G(\varepsilon_g^1)\}+P_{0,1}^{\varsigma_1}\{1-f_G(\varepsilon_g^1)\}+P_{1,0}^{\varsigma_1}\{1-f_L(\varepsilon_s^1)\}=0\nonumber \\
	& -P_{1,0}^{\varsigma_1}\left\{1-f_L(\varepsilon_{s}^1)+f_G(\varepsilon_g^1+U_m)\right\}+P_{1,1}^{\varsigma_1}\left\{1-f_G(\varepsilon_g^1+U_m)\right\}+P_{0,0}^{\varsigma_1}f_L(\varepsilon_s^1)=0\nonumber \\
	&-P_{0,1}^{\varsigma_1}\left\{1-f_G(\varepsilon_{g}^1)+f_L(\varepsilon_s^1+U_m)+\frac{\gamma}{\gamma_c}P^{\varsigma_2}_1\right\}+P_{0,0}^{\varsigma_1}f_G(\varepsilon_g^1)+P_{1,1}^{\varsigma_1}\left\{1-f_L(\varepsilon_s^1+U_m)+\frac{\gamma}{\gamma_c}P^{\varsigma_2}_{0}\right\} = 0\nonumber \\
	& -P_{1,1}^{\varsigma_1}\left\{[1-f_G(\varepsilon_{g}^1+U_m)]+[1-f_L(\varepsilon_s^1+U_m)]+\frac{\gamma}{\gamma_C}P^{\varsigma_2}_0\right\}+P_{1,0}^{\varsigma_1}f_G(\varepsilon_g^1+U_m) +P_{0,1}^{\varsigma_1}\left\{f_L(\varepsilon_s^1+U_m)+\frac{\gamma}{\gamma_c}P^{\varsigma_2}_{1}\right\}=0
	\label{eq:first_sys}
	\end{align} 
\end{widetext}
\begin{align}
& -P_{0}^{\varsigma_2}\{f_R(\varepsilon_s^2)+\frac{\gamma}{\gamma_c}P_{1,1}^{\varsigma_1}\}+P_1^{\varsigma_2}\left\{1-f_R(\varepsilon_{s}^2)+\frac{\gamma}{\gamma_c}P^{\varsigma_1}_{0,1}\right\}=0\nonumber \\
&-P_1^{\varsigma_2}\{1-f_R(\varepsilon_{s}^2)+\frac{\gamma}{\gamma_c}P^{\varsigma_1}_{0,1}\}+P_{0}^{\varsigma_2}\left\{f_R(\varepsilon_s^2)+\frac{\gamma}{\gamma_c}P_{1,1}^{\varsigma_1}\right\}=0,
\label{eq:second_sys}
\end{align}
\normalsize where $\gamma$ and $\gamma_c$ are the associated rates of inter-dot tunneling and   system to reservoir  tunneling  respectively \cite{sispad,dattabook,aniket_nonlocal}. and $f_{\zeta}(\epsilon)$ is  the statistical occupancy probability  of the reservoir $\zeta$ at energy $\epsilon$. For the purpose of our present calculations, we assume quasi-equilibrium electron statistics at the reservoir and hence the function $f_{\zeta}(\epsilon)$ is  the Fermi-Dirac function for the corresponding quasi-Fermi level at reservoir $\zeta$. 
\begin{equation}
f_{\zeta}(\epsilon)=\left\{1+exp\left(\frac{\epsilon-\mu_{\zeta}}{kT_{\zeta}}\right)\right\}^{-1},
\end{equation}
where $T_{\zeta}$ and $\mu_{\zeta}$ are the temperature and quasi-Fermi energy of the reservoir $\zeta$ respectively. The group of Eqns. (\ref{eq:first_sys}) and (\ref{eq:second_sys}) signify  that interdot electron transport between $S_1$ and $S_2$ can only occur when the ground state of $G_1$ is occupied. Both  (\ref{eq:first_sys}) and  (\ref{eq:second_sys}) form dependent sets of equations, which can be broken by employing the probability conservation rules $\sum_{i,j=0,1}P_{i,j}^{\varsigma_1}=1$ and $\sum_{k=0,1}P_{k}^{\varsigma_2}=1$. The sets of Eqns. (\ref{eq:first_sys}) and  (\ref{eq:second_sys}) form a non-linear set of equations and should  be solved using any iterative numerical method. For the purpose of our present calculation, Newton-Raphson scheme was used to calculate the system steady-state probabilities $P_{i,j}^{\varsigma_1}$ and $P_k^{\varsigma_2}$. On calculation of steady-state probabilities the charge current between the system and the reservoirs  $I_{L(R)}$ and the heat current $I_{Qe}$ (extracted from the reservoir $G$) can be calculated as:
\small
\begin{align}
I_L= & q\gamma_c \times \left\{P^{\varsigma_1}_{0,0}f_L(\varepsilon_s^1)+P^{\varsigma_1}_{0,1}f_L(\varepsilon_s^1+U_m)\right\} \nonumber \\ &- q\gamma_c P^{\varsigma_1}_{1,0}\{1-f_L(\varepsilon_s^1)\}- q\gamma_c P^{\varsigma_s^1}_{1,1}\{1-f_L(\varepsilon_s^1+U_m)\}  \\
I_R= & -q\gamma_c \times \left\{P^{\varsigma_2}_{0}f_R(\varepsilon_s^1)-P^{\varsigma_2}_{1}\{1-f_R(\varepsilon_s^1)\}\right\} ,
\label{eq:final}
\end{align}
\begin{equation}
I_{Qe}=U_m \gamma_c\left\{P^{\varsigma_1}_{1,0}f_G(\varepsilon_g+U_m)-P^{\varsigma_1}_{1,1}\{1-f_G(\varepsilon_g+U_m)\}\right\}
\label{eq:heat}
\end{equation}  
\normalsize
\begin{figure}[!htb]
	\includegraphics[width=.4\textwidth]{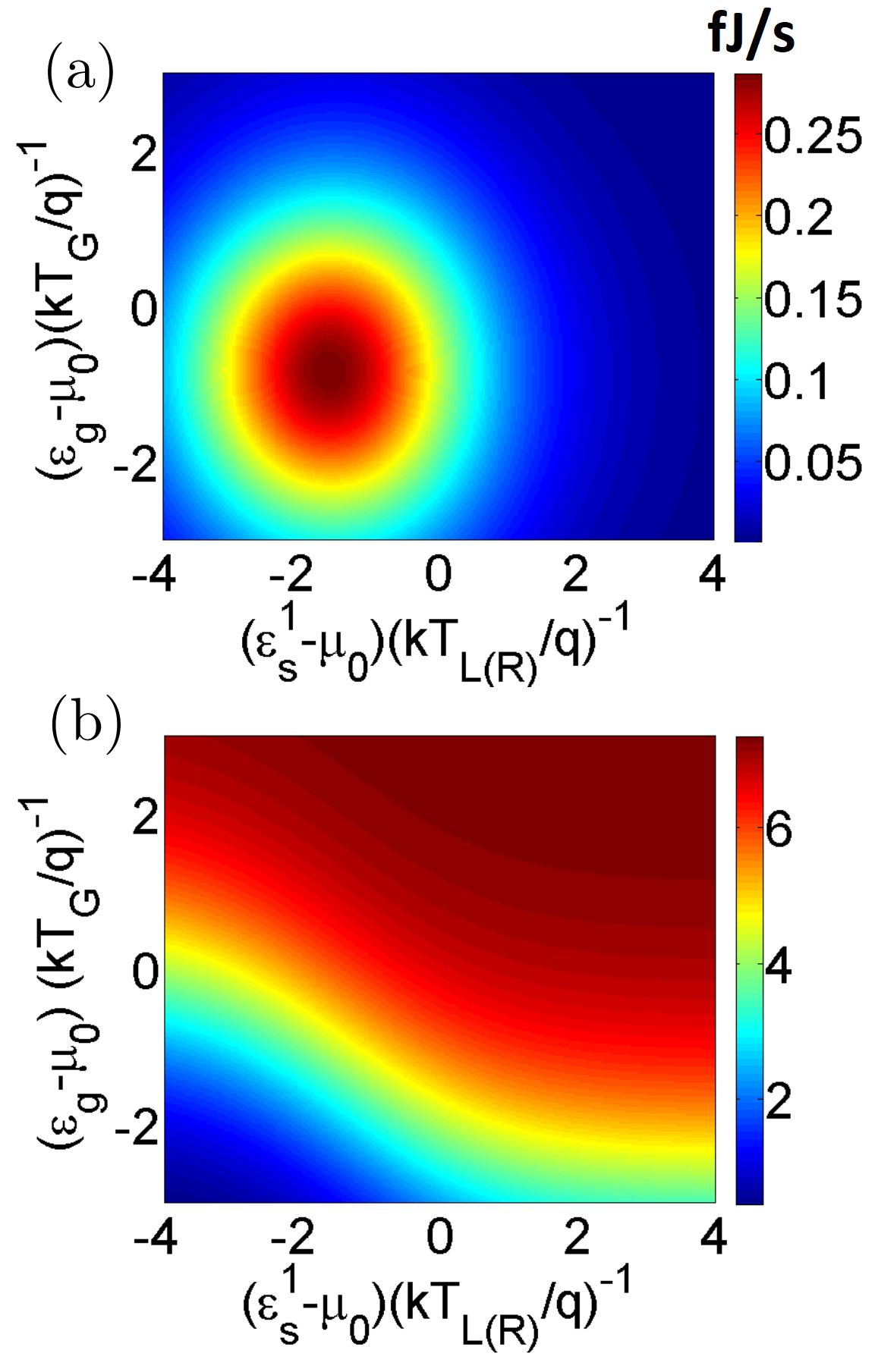}
	\caption{Regime of refrigeration  of the proposed design for low bias condition ($V=0.2\frac{kT}{q}$) with $T_L=T_R=T_G=T=10K$. Color plots are showing the variation of (a) cooling power ($I_{Qe}$) and (b) COP with the position of the ground states $\varepsilon_g$ and $\varepsilon_s^1$ for $V=0.2\frac{kT}{q}~(\approx 0.17meV)$ and $U_m=3\frac{kT}{q}~(\approx 2.5meV)$. }
	\label{fig:Fig_2}
\end{figure}

\begin{figure}[!htb]
	\includegraphics[width=.4\textwidth]{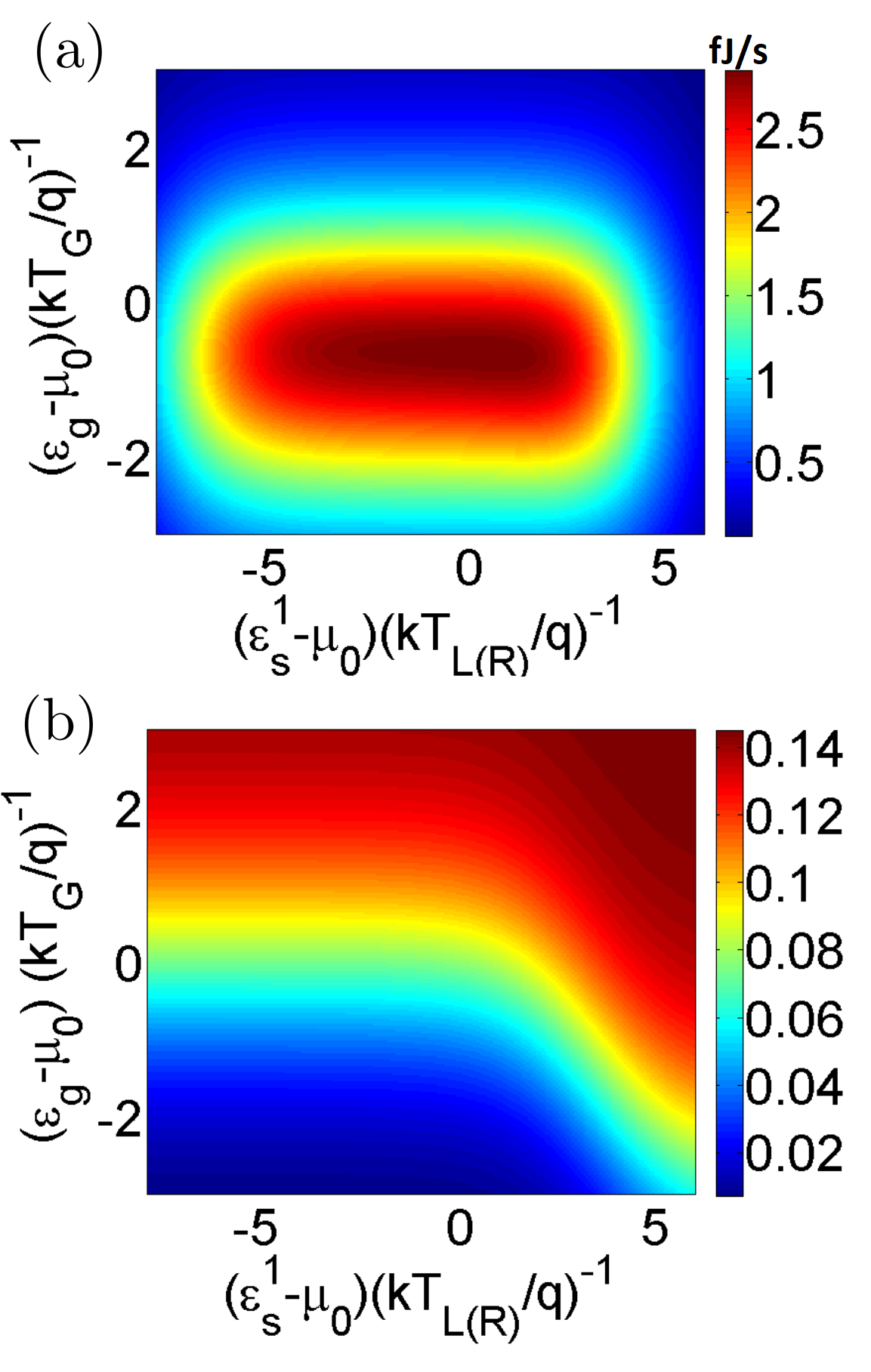}
	\caption{Regime of refrigeration  of the proposed design for high bias condition ($V=10\frac{kT}{q}$) with $T_L=T_R=T_G=T=10K$. Color plots are showing the variation of (a) cooling power ($I_{Qe}$) and (b) COP with the position of the ground states $\varepsilon_g$ and $\varepsilon_s^1$ for $V=10\frac{kT}{q}~(\approx 8.5meV)$ and $U_m=3\frac{kT}{q}~(\approx 2.5meV)$. }
	\label{fig:Fig_3}
\end{figure}

 \begin{figure}[!htb]
	\includegraphics[width=.4\textwidth]{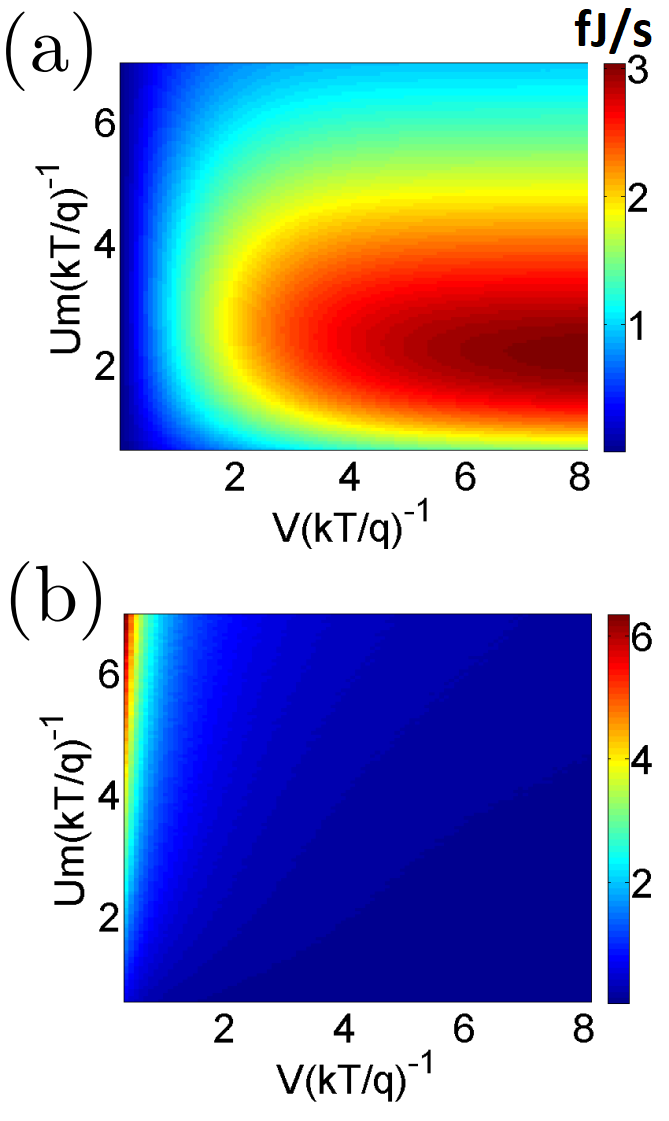}
	\caption{Variation in refrigeration performance of the proposed set-up for  with $T_{L(R)}=10K$ and $T_G=10K$. Color plots are showing the variation of (a) maximum cooling power $I_{Qe}^{M}$ and  (b) $COP$ at the maximum cooling power with the applied bias voltage $V$  and Coulomb coupling energy $U_m$. To find out the  maximum cooling power for a given value of $V$ and $U_m$, the ground states of the dots are tuned to their optimal position. $T=\frac{T_{L(R)}+T_G}{2}$ is the average temperature between the hot and the cold reservoirs.}
	\label{fig:Fig_4}
\end{figure}

\begin{figure}[!htb]
	\includegraphics[width=.47\textwidth]{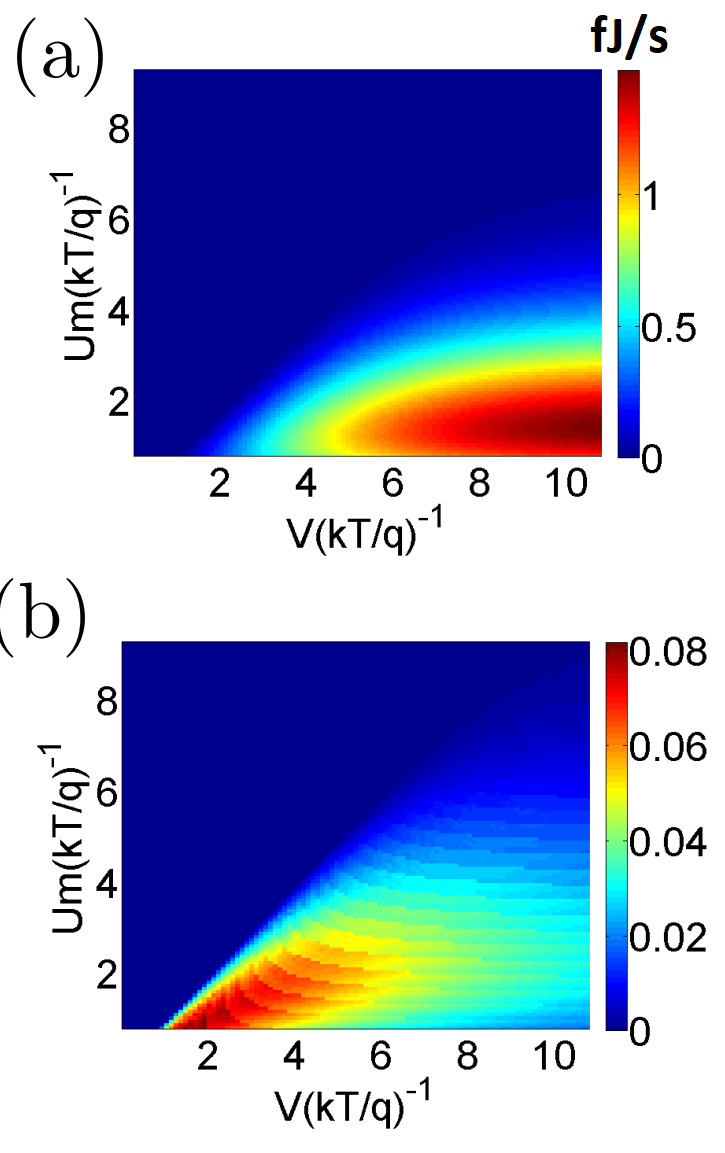}
	\caption{Variation in refrigeration performance of the proposed set-up for  with $T_{L(R)}=10K$ and $T_G=5K$. Color plots are showing the variation of (a) maximum cooling power $I_{Qe}^{M}$ and  (b) $COP$ at the maximum cooling power with the applied bias voltage $V$  and Coulomb coupling energy $U_m$. To find out the  maximum cooling power for a given value of $V$ and $U_m$, the ground states of the dots are tuned to their optimal position. $T=\frac{T_{L(R)}+T_G}{2}$ is the average temperature between the hot and the cold reservoirs.}
	\label{fig:Fig_5}
\end{figure}

\begin{figure}[!htb]
	\includegraphics[width=.5\textwidth]{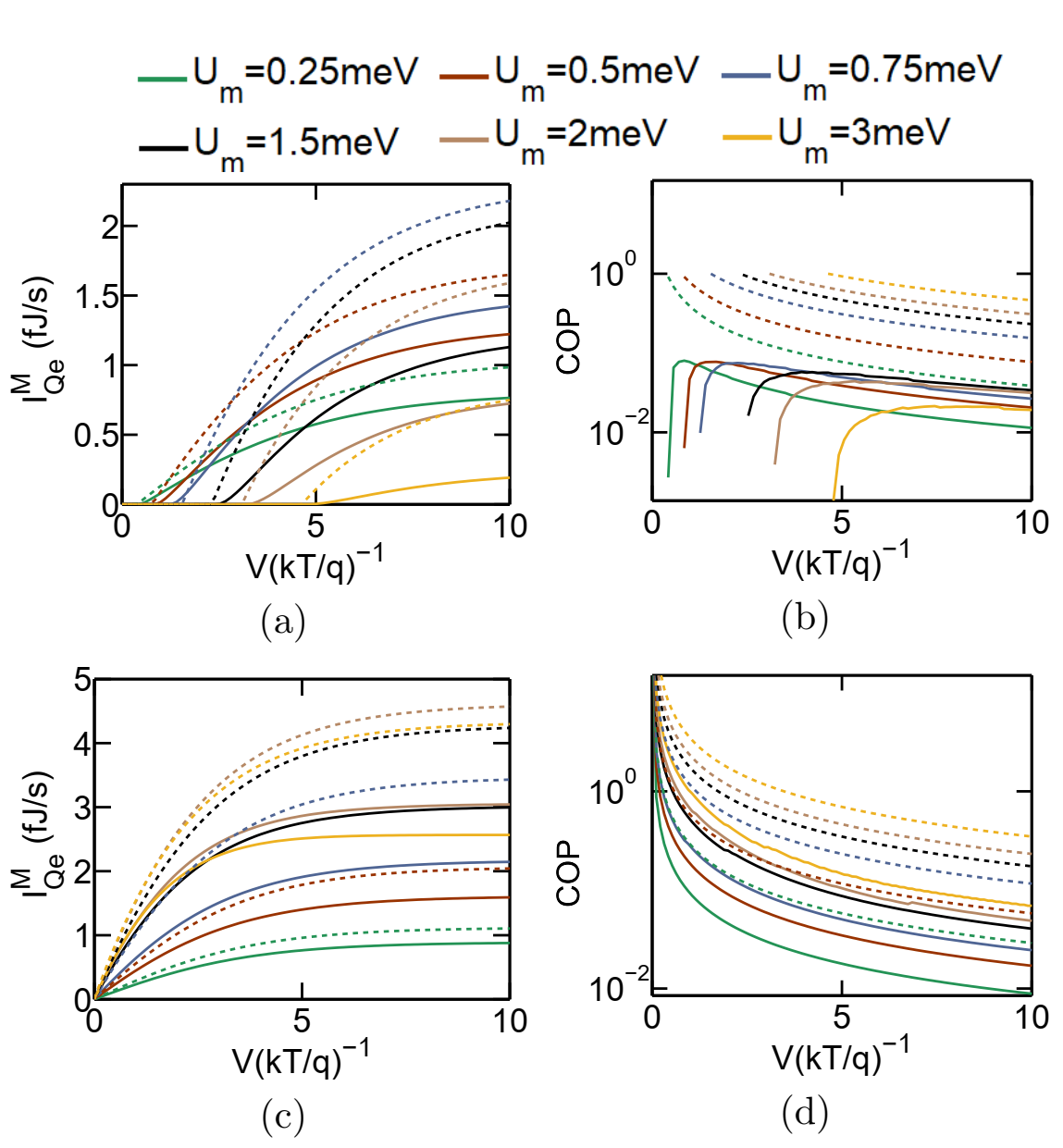}
	\caption{Comparison of performance of the proposed design (solid Lines) with the optimal setup (dashed Lines) for different values of Coulomb coupling energy $U_m$ with $T_L=T_R=10K$ \& $T_G=5K$. Top panel:  (a) Cooling power and  (b) COP (log scale) as a function of bias voltage ($V$)  at $T_{L(R)}=10K,$ and $T_G=5K$. Bottom panel:  (c) Cooling power and  (d) COP (log scale) as a function of bias voltage ($V$) at $T_{L(R)}=10K$, and $T_G=10K$.}
	\label{fig:Fig_6}
\end{figure}

 In Eq.~\eqref{eq:heat}, we have neglected the reverse heat flux due to lattice thermal conductivity, assuming ideal thermal insulation of the reservoir $G$ with its surroundings. It should be noted that, the heat extracted per unit time from reservoir $G$, given in Eq.~\eqref{eq:heat}, is not directly dependent on the ground state $\varepsilon_{g}$ of $G_1$ due to the fact that the net current into (or out of) reservoir $G$ is zero.
As described, to achieve refrigeration in the reservoir $G$, a net electronic flow has to be injected from $L$ to $R$.  To achieve such electron flow, a voltage bias can be applied between $L$ and $R$ with the negative and positive terminals of the bias connected to the terminals $L$ and $R$ respectively. We now briefly discuss the electronic transport processes resulting in the refrigeration of the reservoir $G$. Let us consider that the system is in the initial state $\ket{0,0,0}$, which we also call as the vacuum state. A sequence of electronic transport that extracts a heat packet $U_m$ from the reservoir $G$ is as follows: $\ket{0,0,0}\rightarrow \ket{1,0,0}\rightarrow \ket{1,1,0}\rightarrow \ket{0,1,1}\rightarrow \ket{0,1,0} \rightarrow\ket{0,0,0} $. In this sequence, the system starts with the vacuum state. Next, an electron is injected into $S_1$ with an energy $\varepsilon_{s}^1$ followed by an electron injection in $G_1$ with an energy $\varepsilon_{g}+U_m$. Next, the electron in $S_1$ gets transferred to $S_2$ via interdot tunneling and subsequently flows out of the terminal $R$. The system returns to the initial vacuum state when the electron in $G_1$ tunnels out to $G$ with an energy $\varepsilon_{g}$. Note that in this cycle, the electron is injected into $G_1$ from $G$ with energy $\varepsilon_{g}+U_m$ and extracted back into $G$ with energy $\varepsilon_{g}$. Thus, the reservoir $G$ looses a packet of heat energy $U_m$. Such a  transport process and other equivalent sequential processes lead to a refrigeration in the reservoir $G$. \\
\indent When analyzing the refrigeration performance, two parameters of prime importance constitute the cooling power or the heat extracted per unit time  $I_{Qe}$ (defined in Eq.~\eqref{eq:heat}) and the heat extracted per unit input power, which is also known as the coefficient of performance ($COP$) of the refrigerator. The total input power ($P$) is dependent on the bias voltage as well as the injected current and can be defined as: 
\begin{equation}
P=I_{L(R)} \times V,
\label{eq:pow}
\end{equation}
where $V$ is the applied bias voltage across the reservoir $L$ and $ R$. As stated above,the efficiency of a refrigerator is normally characterized by its coefficient of performance ($COP$):
\begin{equation}
COP=\frac{I_{Qe}}{P},
\end{equation}
where the heat extracted per unit time from $G$ can be calculated using Eq.~\eqref{eq:heat}.
\section{Results}\label{results}
In this section, we describe the operation regime of the proposed refrigeration engine. In addition, we also compare the performance of our design with the optimal non-local refrigerator discussed in literature and elaborate the  transport processes dominating the proposed set-up.  Without loss of generality, we assume that $\gamma_c=10^{-6}\frac{q}{h}$ and $\gamma=10^{-5}\frac{q}{h}$.\\
  \indent \textbf{Analysis of performance and regime of operation:} Fig.~\ref{fig:Fig_2} and \ref{fig:Fig_3} demonstrate the operation regime of the refrigeration engine for low bias ($V=0.2\frac{kT}{q}$) and high bias ($V=10\frac{kT}{q}$) condition respectively for Coulomb coupling energy $U_m=3\frac{kT}{q}$  and  $T_{L(R)}=T_G=T=10K$. 
   In particular, Fig.~\ref{fig:Fig_2}(a) and (b) demonstrate the cooling power $I_{Qe}$ and COP respectively over a range of the ground state positions for $\varepsilon_s^1$ and $\varepsilon_{g}$. It can be noted  from Fig.\ref{fig:Fig_2}(a) that the regime of refrigeration corresponds to
 $\varepsilon_s^1$  lying within a few $kT$ around $\mu_0$, that is $-few~ kT<\varepsilon_s^1-\mu_0<few~ kT$. Such a trend occurs since the rate of electron transport through the system, under low bias condition, peaks when the ground states of the quantum dots lie within a few $kT$ of the equilibrium Fermi-energy $\mu_0$. We also note that the refrigeration power is finite and large when the $\varepsilon_g$ lies within a few $kT$ of the equilibrium Fermi energy $\mu_0$, that is, when  $-few~kT<\varepsilon_g-\mu_0<few~ kT$. Such a behaviour occurs due to the fact that for extraction of heat energy from $G$, an electron must be able to tunnel into and out of $G_1$ with energy $\varepsilon_{g}+U_m$ and $\varepsilon_{g}$ respectively. Hence,  both the functions $f_G(\varepsilon_{g}+U_m)$ and $1-f_G(\varepsilon_{g})$ must have finite values which is only possible if $\varepsilon_{g}$ lies within a $few~kT$ of the equilibrium Fermi-energy $\mu_0$. If $\varepsilon_{g}-\mu_0< - a~few~kT$, then an electrom with energy $\varepsilon_{g}$ wouldn't be able to tunnel out into reservoir $G$. On the other hand,  if $\varepsilon_{g}-\mu_0> ~ a~few~kT$,  there would be no electrons in  $G$ to tunnel into the Coulomb blocked level $\varepsilon_{g}+U_m$. In fact the product $f_G(\varepsilon_{g}+U_m)\{1-f_G(\varepsilon_{g})\}$ is maximized when $\varepsilon_{g}-\mu_0=-\frac{U_m}{2}$. Since $U_m\approx -\frac{3kT}{q}$ in this case, we can note the maximum cooling power occurs around $\varepsilon_{g}-\mu_0=-\frac{3kT}{2q}$.  
Fig.~\ref{fig:Fig_2}  (b) demonstrates the variation in $COP$ for the low bias condition. We note a monotonic increase in COP with $\varepsilon_g$ and $\varepsilon_{s}^1$.  Fig.~\ref{fig:Fig_3} (a) and (b) demonstrates the cooling power and $COP$ for high bias condition with $V=10\frac{kT}{q}$. We note that the refrigeration engine now operates over a wide range of $\varepsilon_{s}^1$, mainly due to the increase in the electron transport window  with an increase in the applied bias $V$. The operation regime, in terms of $\varepsilon_{g}$, however, remains almost the identical to the low bias case. The $COP$, shows a similar trend with the low bias case, that is the $COP$ increases with $\varepsilon_{g}$ and $\varepsilon_s^1$. By comparing  Fig.~\ref{fig:Fig_2} and Fig.~\ref{fig:Fig_3} we note a drastic increase (about $10$ times) in the maximum cooling power. This is due to the fact that an increase in bias voltage causes more electrons to flow between $L$ and $R$, which increases the rate of heat absorption from $G$. The $COP$, on the other hand, decreases drastically with an applied bias voltage. This can be explained by the fact that an increase in the bias voltage causes a higher power dissipation per unit electron flow ($qV$) or per unit heat packet ($U_m$) absorption from $G$, which results in a decrease in $COP$.   It should be noted that an equivalent trend of increase in refrigeration power and decrease in overall COP with increase in bias voltage can also be noted for lower dimensional and bulk Peltier refrigerators \cite{aniket_cool1,aniket_cool2}.  \\
\indent The variation of the optimal performance of the refrigeration engine with variation in the Coulomb coupling energy ($U_m$) and applied bias voltage is demonstrated in Fig.~\ref{fig:Fig_4} and \ref{fig:Fig_5} for the cases $T_G=T_{L(R)}$ and $T_G<T_L(R)$ respectively.  In particular, Fig.~\ref{fig:Fig_4}(a) demonstrates the maximum cooling power ($I_{Qe}^{M}$) and Fig.~\ref{fig:Fig_4}(b) demonstrates the $COP$ at the maximum cooling power for a range of values of the applied bias voltage $V$ and the Coulomb coupling energy ($U_m$). To achieve the maximum cooling power  the ground states of the dots are adjusted to the optimal position, with respect to the equilibrium Fermi-energy. In Fig.\ref{fig:Fig_4}, the maximum cooling power, as well as the $COP$, is low for low values of $U_m$. Despite of a high current for lower values of $U_m$ \cite{sispad}, the total cooling power is low due to low value of the average heat extracted per unit electron flow. Due to low values of heat extracted per unit electron flow, the $COP$ also remains low for low values of $U_m$. The high magnitude of current flow results in a high power dissipation despite of a lower cooling power due to lower values of $U_m$. As $U_m$ increases, the net rate of electron flow  decreases for the same value of bias voltage $V$ \cite{sispad}. However, the average heat extracted per unit electron flow increases with $U_m$. These two competing processes result in an initial  increase in cooling power with an increase in  $U_m$. With further increase in $U_m$ beyond a certain limit, the cooling power finally decreases due to a decrease  in the total current flowing trough the system. From the perspective of the dot $G_1$, it can be stated that the net cooling power decreases with increase in $U_m$ beyond a certain limit due to a lower probability of electrons tunneling into the gate $G_1$ with an energy $\varepsilon_{g}+U_m$, when the ground state of $S_1$ is already occupied. With an increase in the applied bias $V$, we note a monotonic increase in the cooling power, till saturation and a deterioration in the $COP$. The saturation in cooling power at high values of the applied bias ($V> a ~few ~kT/q$) occurs due to a saturation in electronic current through the system.  The reasons of the decrease in $COP$ with an increase in applied bias  has already been discussed with respect to Fig.~\ref{fig:Fig_2} and \ref{fig:Fig_3}. Fig.~\ref{fig:Fig_5}(a) and (b) depicts the refrigeration performance of the system for $T_G<T_{L(R)}$. Here, certain differences should be noted when compared to the refrigeration performance of the system for $T_G=T_{L(R)}$ First of all, the cooling power is non-zero only when the voltage exceeds a certain minimum value, which we call the threshold voltage ($V_{TH}$). The threshold voltage appears due to the presence of a thermoelectric force for $T_G<T_{L(R)}$ which tends to drive electrons from $R$ towards $L$, while dumping heat packets into the reservoir $G$ \cite{aniket_nonlocal}.   Secondly, the voltage beyond which nonzero cooling power is achieved, increases with the increase in Coulomb coupling energy $U_m$. This again is due to the increase in the open-circuited thermoelectric voltage in such a set-up with increase in $U_m$ \cite{aniket_nonlocal}. The applied bias must overcome the thermoelectric voltage to effectively cool the reservoir $G$.  Thirdly, the maximum saturation cooling power, as well as the $COP$ becomes much lower compared to the case of $T_G=T_{L(R)}$. In addition the cooling power at higher values of $U_m$ becomes negligibly small. This again can be explained based on the reverse thermoelectric force acting on the system. Since the applied bias has to inject current against the reverse thermoelectric flux, we get a lower cooling power for a high bias, which, inturn, is responsible for deteriorating the $COP$.  In other words, the regime of operation of the proposed refrigeration engine gets squeezed when $T_G<T_{L(R)}$ (demonstrated in Fig.~\ref{fig:Fig_5}a. An exhaustive discussion and analysis for the case of  $T_G<T_{L(R)}$
 is presented later.\\
  \indent \textbf{Performance comparison with optimal non-local refrigeration engine:} Fig. \ref{fig:Fig_6} demonstrates the performance comparison between the proposed design and optimal non-local refrigeration engine \cite{coulomb_TE1,coulomb_TE2,coulomb_TE3,coulomb_TE4,coulomb_TE5} for two different cases (i) $T_G<T_{L(R)}$ (top panel) and (ii) $T_G=T_{L(R)}$ (bottom panel). For performance comparison, the system to reservoir coupling of the optimal non-local refrigeration engine is taken to be $\gamma_l(\varepsilon)=\gamma_c \theta(\varepsilon_s^1+\delta \epsilon-\varepsilon),~\gamma_l(\varepsilon)=\gamma_c \theta(\varepsilon-\delta \epsilon-\varepsilon_s^1)$ and $\gamma_g(\varepsilon)=\gamma_c$ \cite{coulomb_TE1,coulomb_TE2,coulomb_TE3,coulomb_TE4,coulomb_TE5}, where $\delta \varepsilon <U_m$ and $\theta $ is Heaviside step function. The maximum cooling power ($I_{Qe}^M$) and   $COP$ (in log-scale) as a function of bias voltage $V$ are plotted respectively in the left and right panel of  Fig. \ref{fig:Fig_6} for  $T_G<T_{L(R)}$ (top panel) and  $T_G=T_{L(R)}$ (bottom panel) for different values of $U_m$. For $5K=T_G<T_{L(R)}=10K$, the overall maximum cooling power for the optimal set-up and the proposed design are $2.1J/s$ and $1.45J/s$ respectively. The overall maximum cooling power, in both the set-ups is achieved at $U_m=0.75meV~(\approx 1.75\frac{kT_G}{q})$. Similarly, for $T_G=T_{L(R)}=10K$, the overall maximum cooling power for the optimal set-up and the proposed set-up are $4.6J/s$ and $3J/s$ respectively. In this case, the maximum cooling power for both the set-ups occur at $U_m=2meV ~(\approx 2.3\frac{kT_G}{q})$. Thus, in both the cases, the overall maximum cooling power of the proposed design hovers around $65-70\%$ of that of the optimal set-up. Fig.~\ref{fig:Fig_6} (b) and (d) depicts the $COP$ (log-scale) for the proposed set-up (solid lines) and the optimal design (dashed lines) for the cases $T_G<T_{L(R)}$ and $T_G=T_{L(R)}$ respectively. We note that the $COP$ for the  proposed design is much less than that of the optimal set-up. This is because in our proposed set-up a fraction of the total number of electrons flow from $L$ to $R$ without absorbing heat from the reservoir $G$ (explained in the next part).\\
  \indent \textbf{Sequential tunneling mechanism leading to a performance degradation:} 
  \begin{figure}[!htb]
  	\includegraphics[width=.5\textwidth]{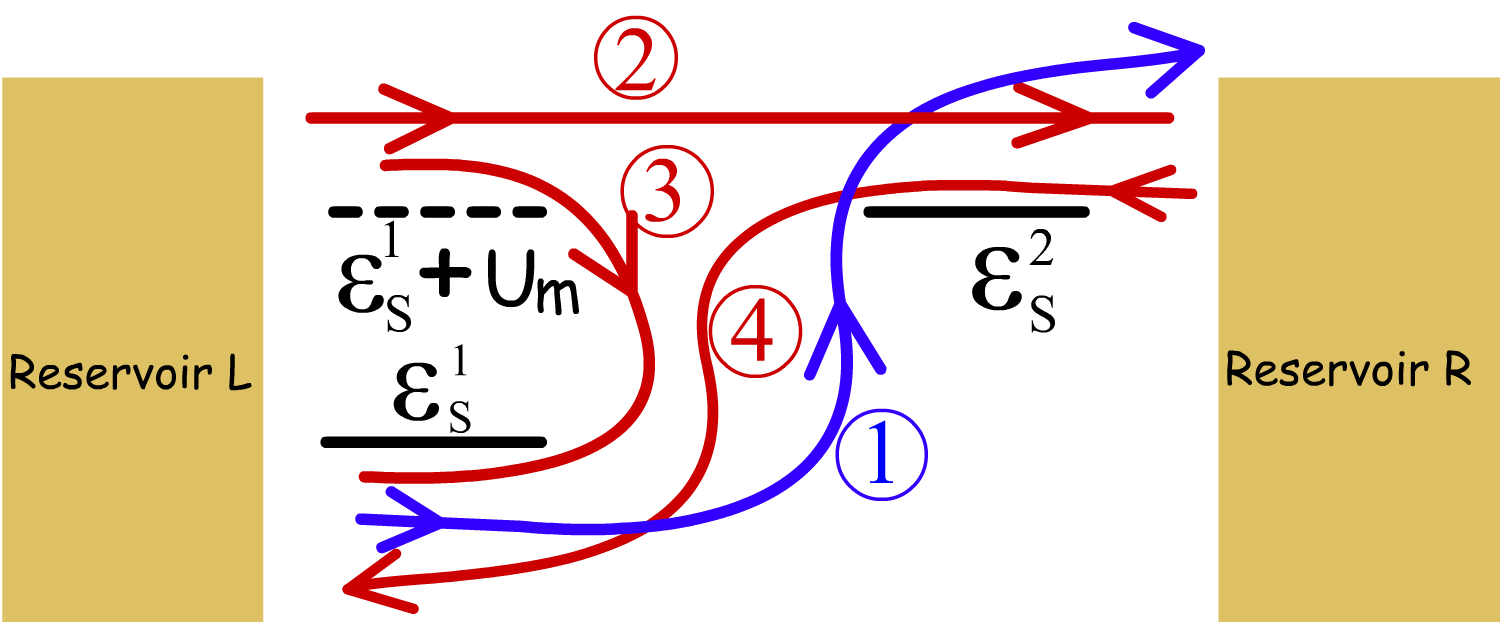}
  	\caption{Schematic diagram demonstrating the different electronic current components  from the reservoir $L$ to the system through the energy level $\varepsilon_{s}^1$ and the Coulomb blockaded level $\varepsilon_s^1+U_m$. Four current components are shown in the figure. $(1)$ electron current  flows due to voltage bias from the reservoir $L$ to the $R$ while absorbing a heat packet $U_m$ (per electron) from $G$. (2) Electron current, due to applied bias, flows directly from $L$ to $R$ without any heat absorption.  (3) and (4): Electron current flow due to thermoelectric force, that tends to flow while dumping heat packet $U_m$ into the reservoir $G$.}
  	\label{fig:Fig_7}
  \end{figure}
  \begin{figure}[!htb]
  	\includegraphics[width=.45\textwidth]{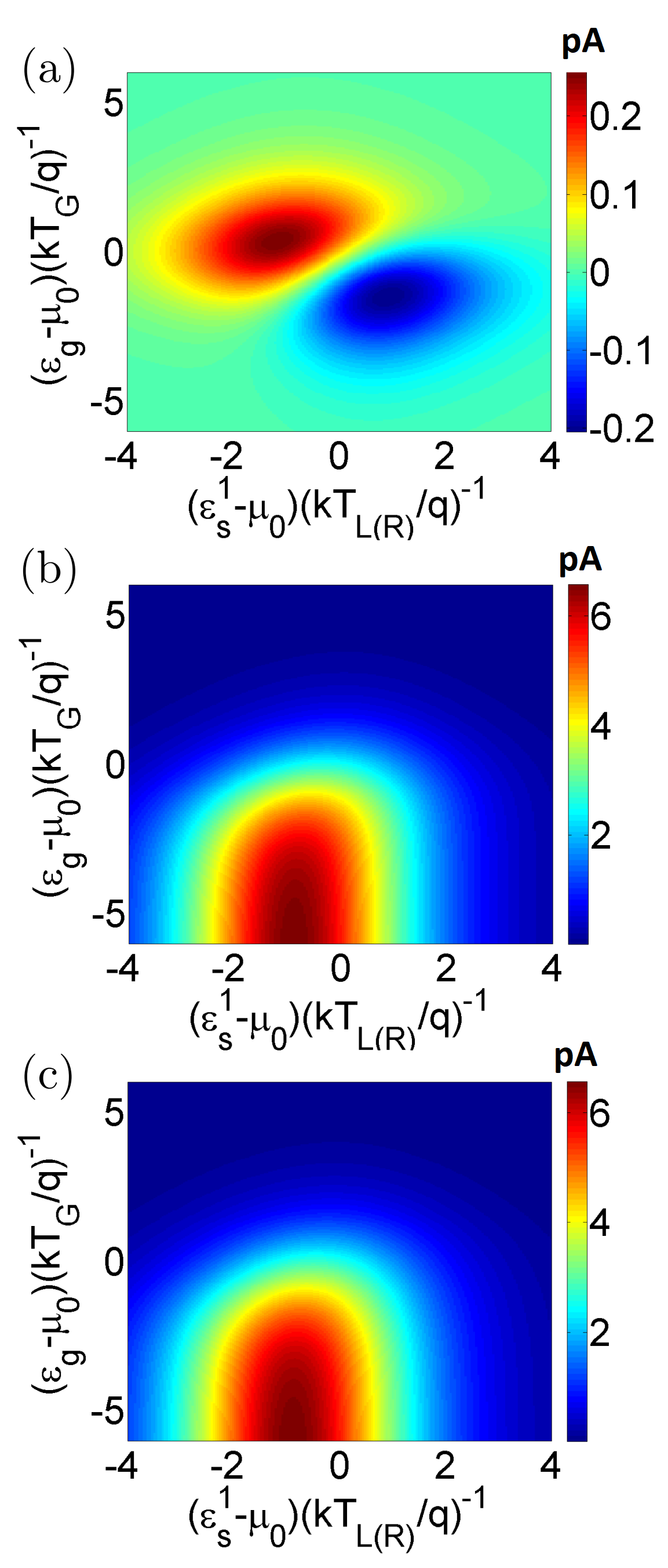}
  	\caption{Colour plots demonstrating the  electron flow into the system from the reservoir $L$ with variation in the the ground states $\varepsilon_g$ and $\varepsilon_s^1$, for $U_m=0.75meV~(\approx 1.16\frac{kT}{q})$ and  $V=1.3meV~(\approx 2\frac{kT}{q})$, through (a)  the ground state $\varepsilon_{s}^1$ (b) the Coulomb blockaded level $\varepsilon_{s}^1+U_m$. (c) Total  average electron current between the system and the reservoir $L$. }
  	\label{fig:Fig_8}
  \end{figure}
 \indent Now, we discuss the sequential transport mechanisms leading to a performance degradation of the proposed refrigeration engine. Let us consider the sequence of  electron transport from  $L$ to  $R$ that results in absorbing a heat packet $U_m$ from reservoir $G$.  For example, in the sequence $\ket{0,0,0}\rightarrow \ket{1,0,0}\rightarrow\ket{1,1,0}\rightarrow\ket{0,1,1}\rightarrow\ket{0,1,0}\rightarrow\ket{0,0,0}$, the system initially starts with the vacuum state $\ket{0,0,0}$. An electron is injected from reservoir  $L$ into $S_1$ at energy $\varepsilon_{s}^1$, followed by another electron injected  into $G_1$ from $G$ with an  energy $\varepsilon_{g}+U_m$. This is followed by, the electron in $S_1$ tunneling into $S_2$, after which the electron present in $G_1$ tunnels out into reservoir $G$ with an energy $\varepsilon_g$. At the end of the cycle, the system returns to the vacuum state when the electron present in $S_2$ tunnels out, with  energy $\varepsilon_s^2=\varepsilon_{s}^1+U_m$, into $R$. Thus, in the entire  process illustrated above an electron is transmitted from the reservoir $L$ to $R$ while absorbing a heat packet $U_m$ from $G$. These type of transport processes contribute to refrigeration of the target reservoir $G$ and are illustrated as (1)  in Fig.~\ref{fig:Fig_7}.
 The second electron transport component, depicted in Fig.~\ref{fig:Fig_7} as (2), results in direct transmission of electrons between $L$ and $R$ without an absorption of heat packets from $G$. Hence, this component  results in wastage of power, thereby causing a degradation of $COP$.
  Next, let us consider the following sequence of electron transport $\ket{0,0,0}\rightarrow \ket{1,0,0}\rightarrow\ket{1,1,0}\rightarrow\ket{0,1,0}\rightarrow\ket{0,0,0}$. In the above sequence, an electron tunnels, with an energy $\varepsilon_{s}^1$, into $S_1$ from reservoir $L$. This is  followed by an electron entering $G_1$, from reservoir $G$, with an energy $\varepsilon_g+U_m$. At the next step, the electron present in $S_1$ tunnels out into reservoir $L$ with an energy $\varepsilon_{s}^1+U_m$. At the end of the sequence, the electron in $G_1$ exits into reservoir $G$ with an energy $\varepsilon_{g}$. It is evident that in this process, a packet of heat energy $U_m$ is transmitted from reservoir $L$ to $G$. Thus, this kind of sequence results in heating up of the target reservoir $G$ and is only positive and finite for $T_G<T_{L(R)}$ \cite{aniket_nonlocal}. This current component, depicted in Fig.~\ref{fig:Fig_7} as (3), flows due to thermoelectric force and affects the refrigeration performance of the proposed refrigeration engine by transmitting heat packets into $G$. Another thermoelectric  component, that flows for $T_G<T_{L(R)}$, while dumping heat packets into $G$ is  shown as (4) in Fig.~\ref{fig:Fig_7}. From Fig.~\ref{fig:Fig_7} and the above discussions, it is clear that the electron flow components (2) and (3) from $L$ into the  Coulomb blockaded level $\varepsilon_{s}^1+U_m$ result in degradation of the refrigeration performance.  \\
 \indent To further elaborate the above discussion, we separate out the current flow into the system from the reservoir $L$ as:
 \[
 I_L=I_{1}+I_{2},
 \]
 where 
 \begin{align}
 &I_{1}=q\gamma_c \times \left\{P^{\varsigma_1}_{0,0}f_L(\varepsilon_s^1)-P^{\varsigma_1}_{1,0}(1-f_L(\varepsilon_s^1)\right)\} \nonumber \\ 
 &I_{2}= q\gamma_c \{P^{\varsigma_1}_{0,1}f_L(\varepsilon_s^1+U_m)-  P^{\varsigma^1}_{1,1}\{1-f_L(\varepsilon_s^1+U_m)\}\} 
 \label{eq:split}
 \end{align}
 In the above equation, $I_{1}$ and $I_{2}$ denote the total electron current from  reservoir $L$ to  the energy level $\varepsilon_s^1$ and  the Coulomb blockaded  level $\varepsilon_s^1+U_m$ respectively.  Particularly, Fig.~\ref{fig:Fig_8}(a) and (b) respectively  depict the electron current flow into the system from reservoir $L$ via the energy level $\varepsilon_s^1$ ($I_{1}$) and  the Coulomb blockaded  level $\varepsilon_s^1+U_m$ ($I_{2}$).  Fig.~\ref{fig:Fig_8}(c), on the other hand, depicts the overall electronic flow $I_L=I_{1}+I_{2}$ from the reservoir $L$ into the system.

  We find that  the electron current flow $I_{1}$ through $\varepsilon_{s}^1$ into the system from $L$ is positive, over certain regime and negative over the rest. The positive regime corresponds to refrigeration of the target reservoir $G$. The negative regime, on the other hand, has the potential for thermoelectric generation where the flux of electrons flow against the voltage bias due to thermoelectric force \cite{aniket_nonlocal}. The regime with negative value of $I_1$ corresponds to no net cooling of the target reservoir $G$.  Interestingly, we also find that the current component $I_{2}$ through the Coulomb blockaded energy level $\varepsilon_{s}^1+U_m$ is positive, as already shown in Fig.~\ref{fig:Fig_8}(b). This electron current constitutes the components (2) and (3). As already discussed, these current components results only in deterioration of the net cooling power, as well as, the $COP$.   Thus, they negatively affect   the refrigeration performance.  The   deterioration in  performance of the refrigeration engine, due to these current components (2) and (3), which enter the system via the Coulomb blockaded level $\varepsilon_{s}^1+U_m$, can be chocked by adding an extra filter. Such an electron filter is to be added between $L$ and $S_1$ to restrict electron flow via the Coulomb blocked level $\varepsilon_s^1+U_m$ to reduce  the current components (2) and (3). However, doing so neutralizes the novelty of the proposed set-up in terms of fabrication simplicity. In Fig.~\ref{fig:Fig_8}(c), we show the total electronic current flow from $L$ to $S_1$. It is clear that a very large portion of the total  electron flow actually consists of the component (2), which results in a lower $COP$ in the proposed system compared to the optimal design. \\
  
 \begin{figure}[!htb]
 	\includegraphics[width=.4\textwidth]{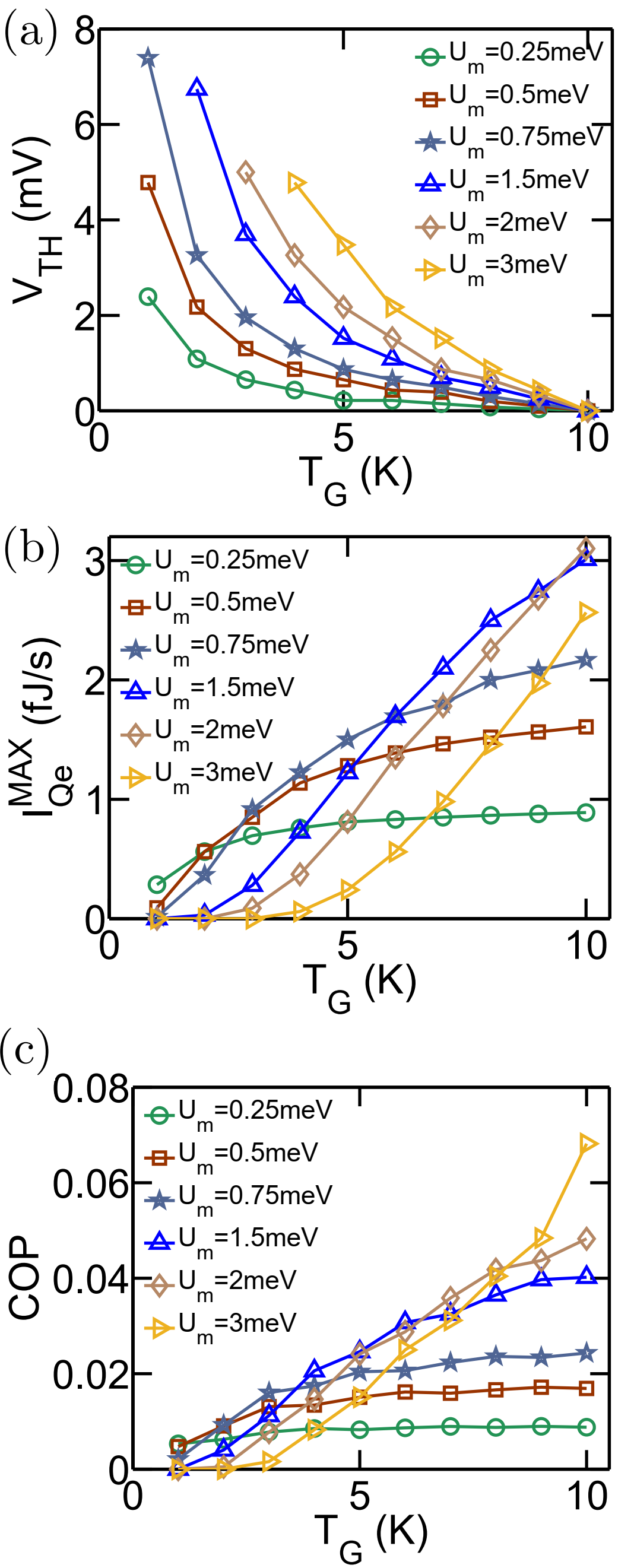}
 	\caption{Performance variation of the proposed refrigeration engine as the target temperature of the reservoir $G$ ($T_G$) is gradually decreased beyond $T_{L(R)}$. Variation in (a) Threshold voltage $V_{TH}$ (b) overall maximum cooling power $I_{Qe}^{MAX}$ and (c) COP at the maximum cooling power (for $V=10\frac{kT_{L(R)}}{q}$) with decrease in the target temperature $T_G$ of the reservoir $G$.}
 	\label{fig:Fig_9}
  \end{figure}

\indent \textbf{Performance analysis for $\boldsymbol{T_G<T_{L(R)}:}$} For a practical electronic refrigeration engine, the target temperature of the reservoir $G$ should generally be less than the environmental temperature or the average temperature of the current path. We have already noted that when the temperature of the target reservoir $G$ is less than $T_{L(R)}$, the applied bias voltage needs to be greater than a certain threshold voltage $V_{TH}$ due to reverse thermoelectric flux \cite{aniket_nonlocal}.  Fig.~\ref{fig:Fig_9} demonstrates the performance of the proposed refrigeration engine as the temperature of reservoir $G$ is gradually reduced below $T_{L(R)}$. Specifically, Fig.~\ref{fig:Fig_9}(a) depicts the variation in the required threshold voltage to achieve refrigeration, while Fig.~\ref{fig:Fig_9}(b) demonstrates the variation in maximum (saturation) cooling power $I_{Qe}^{MAX}$ with $T_G$ for different values of $U_m$. From Fig. \ref{fig:Fig_9}, we can infer a number of points on the practical conditions of operations of the refrigeration engine, depending on the target temperature of the reservoir $G$.  We note from Fig.~\ref{fig:Fig_9}(a), that the threshold voltage $V_{TH}$ is zero for $T_G=T_{L(R)}$ and increases with decrease in $T_G$. This is because the non-local thermoelectric voltage, which acts opposite to the bias voltage increases with increase in  $\Delta T=T_{L(R)}-T_G$. This basically means that to achieve a lower target temperature $T_G$, even under ideal conditions of \textit{zero} lattice thermal conductivity, we need to apply a higher voltage bias $V$. We also note that the threshold voltage $V_{TH}$ increases with increase in $U_m$. This is due to an increase in the open-circuit thermoelectric voltage with an increase in $U_m$ \cite{aniket_nonlocal}. Fig.~\ref{fig:Fig_9}(b) demonstrates the variation in the saturation cooling power (high bias voltage limit) with a decrease in the target temperature $T_G$ for various values of $U_m$. With a decrease in the target temperature the saturation cooling power decreases monotonically. An interesting point to note is that for higher values of target temperature $T_G$, the saturation cooling power is higher for higher values of $U_m$. On the other hand, when one approaches smaller target temperature $T_G$, the saturation cooling power is higher for lower values of $U_m$. In particular, we note that beyond $T_G\leq 2K$, the saturation cooling power for $U_m\geq 1.5meV$ is approximately \textit{zero}, which basically points out that a system with $U_m \geq 1.5meV$ cannot be employed to achieve a temperature beyond $2K$, even under ideal conditions. This can be understood by the fact that a non-zero cooling power demands finite value for both $f_G(\varepsilon_{g}+U_m)$ and $1-f_G(\varepsilon_{g})$, so that electrons can tunnel into $G_1$ with energy $\varepsilon_{g}+U_m$ and tunnel out into reservoir $G$ with energy $\varepsilon_{g}$. At very low temperature, the smearing of Fermi function $f_G$ around $\mu_0$ decreases significantly, making the product $f_G(\varepsilon_{g}+U_m)\{1-f_G(\varepsilon_{g})\}\approx 0$ for any $\varepsilon_{g}$ at higher values of $U_m$. Hence, at lower target temperature $T_G$, the saturation or maximum value of cooling power becomes approximately zero  for higher values of $U_m$. Thus for higher values of $U_m$, even by applying a high bias voltage, we cannot achieve a target temperature $T_G$ lower that a certain minimum temperature (since the saturation cooling power becomes zero beyond a certain minimum temperature). We hence conclude that even under ideal conditions, depending on the Coulomb coupling energy $U_m$, there is a minimum limit beyond which the target temperature $T_G$ cannot be reduced. So,  for achieving a lower target temperature $T_G$, one needs to design a system with lower value of $U_m$ and  operate it at a higher value of bias voltage $V$. Fig.~\ref{fig:Fig_9}(c) demonstrates the variation of  $COP$ with target reservoir temperature at the maximum cooling power for $V=\frac{10kT}{q}$. We note that there is fall in $COP$ as one approaches lower target reservoir temperature, the fall being  more sharper for higher values of $U_m$. This is because a higher magnitude of $U_m$ results in a higher value of the open-circuit thermoelectric voltage as one approaches lower values of target reservoir temperature. A higher thermoelectric voltage, acting opposite to the bias voltage, results in a sharper decrease in $COP$ with fall in the target reservoir temperature for higher values of $U_m$.  Although not demonstrated in this paper, in the case of $T_G>T_{L(R)}$, the target reservoir $G$ is automatically cooled (without an applied bias voltage) with $T_G$ gradually approaching $T_{L(R)}$, due to the presence of a thermoelectric force that tends to drive a current between $L$ and $R$ while extracting heat from reservoir $G$ \cite{aniket_nonlocal}.


\section{Conclusion}\label{conclusion}

In this paper, we have proposed a realistic design for non-local refrigeration engine based on Coulomb coupled systems.  The performance of the proposed refrigeration engine was then theoretically investigated employing the quantum master equation (QME) formalism. It was demonstrated that the maximum cooling power of the proposed set-up hovers around $65-70\%$ of the optimal design proposed in literature \cite{coulomb_TE1,coulomb_TE2,coulomb_TE3,coulomb_TE4,coulomb_TE5}. Despite a lower cooling power, the key edge of the proposed set-up over the optimal design is the integration of fabrication simplicity along with decent refrigeration performance. In our discussions, we have restricted
transport phenomena in the weak coupling regime, so that  co-tunneling processes can be
neglected. The refrigeration  power in the proposed set-up
can be enhanced by a few orders  by tuning
electronic transport in the regime of strong coupling, that
is by enhancing the system-to-reservoir and the
interdot tunnel coupling. An analysis on 
the effects of cotunneling and higher order processes on the refrigeration performance
of the proposed system constitutes an interesting aspect of investigation. Furthermore, an analysis of the impact of electron-phonon
scattering \cite{phonon1,phonon2,phonon3,phonon4,phonon5,phonon6,phonon7,superlattice1,superlattice2,nanoflake_heat,nanowire_heat1,nanowire_heat2} on the performance of the  proposed refrigeration engine also constitutes an
interesting aspect of future research. The various 
possible designs  for non-local refrigeration systems  is left for future investigation.  Nevertheless, the design
demonstrated  here can be used to realize high
performance non-local cryogenic refrigeration engines employing Coulomb coupled systems.

\bibliography{apssamp}
\end{document}